\documentclass[prd, superscriptaddress, tightenlines, longbibliography, nofootinbib, eqsecnum, amsfonts, amsmath, floatfix, twocolumn, notitlepage]{revtex4-2}
\pdfoutput=1
\usepackage{ftnxtra}
\usepackage[utf8]{inputenc}
\usepackage{mathrsfs}
\usepackage{euscript}
\usepackage{graphics}
\usepackage{graphicx}
\usepackage{amsmath}
\usepackage{amssymb}
\usepackage{gensymb}
\usepackage{tabularx}
\usepackage{subfigure}
\usepackage{bm}
\usepackage[usenames,dvipsnames,svgnames,table]{xcolor}
\definecolor{linkcolor}{rgb}{0.6,0,0}
\definecolor{citecolor}{rgb}{0,0,0.75}
\definecolor{urlcolor}{rgb}{0.12,0.46,0.7}
\usepackage{xspace}
\usepackage{wasysym}
\usepackage{times}
\usepackage{appendix}
\usepackage{comment}
\usepackage{lipsum}
\usepackage[nolist,nohyperlinks]{acronym}
\usepackage{float}
\usepackage{simplewick}
\usepackage{natbib, ifthen}
\usepackage[breaklinks, colorlinks, urlcolor=urlcolor, linkcolor=linkcolor,citecolor=citecolor,pdfencoding=auto]{hyperref}
\usepackage{listings}
\usepackage{caption}

\newcommand{\mksym}[1]{\ifmmode {\rm #1}\else #1\fi}

\providecommand{\text}[1]{\rm{#1}}

\renewcommand{\d}{\text{d}}
\newcommand{\grad}{\nabla}

\newcommand{\begm}{\begin{pmatrix}}
\newcommand{\enm}{\end{pmatrix}}

\newcommand\ba{\begin{eqnarray}}
\newcommand\ea{\end{eqnarray}}
\newcommand\bea{\begin{eqnarray}}
\newcommand\eea{\end{eqnarray}}

\newcommand\be{\begin{equation}}
\newcommand\ee{\end{equation}}

\newcommand{\valpha}{{\boldsymbol{\alpha}}}

\newcommand{\vell}{{\boldsymbol{\ell}}}

\newcommand{\boldvec}[1]{{\mbox{\boldmath{$#1$}}}}

\newcommand{\vL}{\boldvec{L}}

\newcommand{\vn}{\boldvec{n}}

\defcitealias{PL2018}{PL2018}

\defcitealias{Carron:2022eyg}{PL4}

\newcommand{\Cov}[0]{ {\rm{Cov}} }

\newcommand{\diff}{\mathop{}\!\mathrm{d}}

\newcommand{\av}[1]{\left \langle #1\right\rangle}

\begin{document}

\title{Cluster profiles from  beyond-the-QE CMB lensing mass maps}

\newcommand{\Geneve}{Universit\'e de Gen\`eve, D\'epartement de Physique Th\'eorique et CAP, 24 Quai Ansermet, CH-1211 Gen\`eve 4, Switzerland}
\newcommand{\IISER}{Department of Physics, Indian Institute of Science Education and Research, Pune 411008, India}
\newcommand{\RRI}{Raman Research Institute, C. V. Raman Avenue, Sadashivanagar, Bengaluru 560080, India}
\newcommand{\ICTP}{Instituto de F\'isica Te\'orica da Universidade Estadual Paulista and ICTP South American Institute for Fundamental Research,
R. Dr. Bento Teobaldo Ferraz, 271, Bloco II, Barra-Funda - S\~ao Paulo/SP, Brasil}

\author{Sayan Saha}
\email{sayan.saha@students.iiserpune.ac.in}
\affiliation{\Geneve}
\affiliation{\IISER}
\affiliation{\RRI}
\author{Louis Legrand}
\email{louis.legrand@unige.ch}
\affiliation{\Geneve}
\affiliation{\ICTP}
\author{Julien Carron}
\email{julien.carron@unige.ch}
\affiliation{\Geneve}

  \begin{abstract}
 Clusters of galaxies, being the largest collapsed structures in the universe, offer valuable insights into the nature of cosmic evolution. Precise calibration of the mass of clusters can be obtained by extracting their gravitational lensing signal on the Cosmic Microwave Background (CMB) fluctuations. We extend and test here the performance achieved  on cluster scales by the parameter-free, maximum a posteriori (MAP) CMB lensing reconstruction method, which has been shown to be optimal in the broader context of CMB lensing mass map and power spectrum estimation.  In the context of cluster lensing, the lensing signal of other large-scale structures acts as an additional source of noise.  We show here that by delensing the CMB fluctuations around each and every cluster, this noise variance is reduced according to expectations. We also demonstrate that the well-known bias in the temperature quadratic estimator in this regime, sourced by the strong non-Gaussianity of the signal, is almost entirely mitigated without any scale cuts. Being statistically speaking an optimal and blind lensing mass map reconstruction, the MAP estimator is a promising tool for the calibration of the masses of clusters.
  \end{abstract}


   \maketitle

\section{Introduction}
\setcounter{footnote}{0}

Galaxy clusters are the largest gravitationally bound structures in the Universe. Their abundance as a function of redshift and mass is a direct probe of the growth of structures, and thus provides constraints on the matter density $\Omega_{\rm m}$, on the amplitude of matter fluctuations $\sigma_8$, as well as on the dark energy equation of state and the sum of neutrino masses \cite{Vikhlinin:2008ym,Sehgal:2010ca,Allen:2011zs,Planck:2013lkt, Mantz:2014xba,Mantz:2014paa, Planck:2015lwi,SPT:2016izt, SPT:2018njh, SPT:2021efh, Raghunathan:2021zfi}.

To constrain the cluster mass, most surveys rely on either the richness \cite[e.g.][]{Koester:2007bj,DES:2015mqu,Andreon:2016eck, Farahi:2016xux,Simet:2016mzg}, the X-Ray emission \cite[e.g.][]{Arnaud:2005ur, Arnaud:2007br, Vikhlinin:2008ym}, or the Sunyaev-Zel'dovich (SZ) effect \cite[e.g.][]{Vanderlinde:2010eb, Planck:2013lkt,Planck:2015lwi}. These observables need to be calibrated against the total cluster mass (including both baryonic and dark matter).
This scaling relation between the observable and the cluster mass is currently the dominant source of systematic error \cite{Pratt:2019cnf, Salvati:2020exw, Salvati:2021gkt}.
With future CMB surveys, which are expected to detect of order $10^5$ galaxy clusters \cite{Madhavacheril:2017onh, SimonsObservatory:2018koc, CMB-S4:2016ple, Raghunathan:2021zfi}, the systematic uncertainties will dominate the error budget, and as such accurate mass calibration is even more necessary.

The effect of gravitational lensing provides a direct observable of the total mass, free from assumptions on the dynamical state of the gas.
Gravitational lensing of background galaxies offers precise calibration of the scaling relation \cite{vonderLinden:2014haa, Hoekstra:2015gda, Smith:2015qhs, Sereno:2017zcn, Penna-Lima:2016tvo, Bellagamba:2018gec,Miyatake:2018lpb, Umetsu:2020wlf}, but is subject to systematics such as the intrinsic alignment and redshift uncertainties of the sources \cite{Becker:2010xj}, and is limited by the absence of background galaxies for clusters at high redshift.

The gravitational lensing of the CMB, on which we focus here, is free from the systematics of galaxy weak lensing, and most useful for high redshift clusters. 
The CMB acts as an extended source at a redshift of 1100, with well understood statistics \cite{Lewis:2006fu}.
Note that CMB is not free from systematics, mainly due to astrophysical contaminations, such as the own SZ emission of the cluster \cite{Madhavacheril:2018bxi, DES:2018myw, Patil_2020}.
However, at the small scales relevant for CMB lensing, polarized foreground emission is expected to be negligible.

The standard tool to reconstruct the lensing deflection field from the CMB is the quadratic estimator (QE) \cite{Hu:2001tn, Hu:2001kj, Okamoto:2003zw, Planck:2018lbu}, which combines pairs of CMB maps to estimate the statistical anisotropies created by lensing.
It has been shown that the standard temperature QE underestimates the strong deflection field of massive clusters due to a bias in the estimated gradient (unlensed) CMB field \cite{Maturi:2004zj}. Filtering out the small scales on the gradient leg of the QE greatly reduces this bias \cite{Hu:2007bt}, at the expense of losing some moderate amount of signal to noise.

For current and upcoming CMB surveys, the QE reconstructed cluster lensing field is noise dominated. One possibility is to stack the reconstructed lensing map and measure the average cluster mass \cite{DES:2017fyz, Geach:2017crt, DES:2018myw, ACT:2020izl}. Similarly, one can obtain the mean cluster mass with matched-filtering \cite{Melin:2014uaa, Louis:2016gvv, Zubeldia:2019brr, Zubeldia:2020knz}.

Other estimators have been developed to reconstruct the lensing field around galaxy clusters, such as stacking CMB maps along the gradient direction to extract the lensing dipole \cite{SPT:2019qkp, Levy:2023moy}, the gradient inversion estimator \cite{Horowitz:2017iql, Hadzhiyska:2019cle}, or the maximum likelihood estimator (MLE) \cite{Lewis:2005fq,Baxter:2014frs, Raghunathan:2017cle}.
The MLE directly fits the few parameters of a lensing template on observed CMB maps, and thus estimates the cluster mass without reconstructing the actual lensing map signal. This estimator is adequate for stacking analyses, provided the template shape accurately describes the mass profile and the mass distribution around the cluster, including potential sources of contamination such as the SZ effect.
Recently, machine learning also provided a tool to estimate the cluster mass from lensed CMB maps, using neural networks trained on simulations \cite{Caldeira:2018ojb, Gupta:2020him, Parker:2022uxh}.

First introduced in \cite{Hirata:2002jy, Hirata:2003ka}, likelihood-based CMB lensing mass map estimators aim to optimally reconstruct the large scale deflection field.
Refs.~\cite{ Yoo:2008bf, Yoo:2010jd} introduced similar technology in the context of cluster lensing.
Their approach is built on assuming a cluster convergence profile, and iteratively delensing the observed maps to obtain the cluster mass. The lensing template at each iteration is estimated by stacking a set of reconstructed cluster convergence profiles. Some approximations are made for a tractable implementation, due to issues sourced by the beam of the instrument.

Recent years saw further development of the CMB lensing likelihood-based estimator of Ref.~\cite{Hirata:2002jy, Hirata:2003ka} (often called now maximum a posteriori estimator (MAP)), and other Bayesian attempts, \cite{Carron:2017mqf,Millea:2017fyd,Millea:2020cpw, Millea:2021had,Legrand:2021qdu,Aurlien:2022tlp,Legrand:2023jne,Reinecke:2023gtp}.
It has been demonstrated that the MAP estimator outperforms the QE, in particular for deep polarization surveys such as CMB-S4 \cite{CMB-S4:2016ple}, where most of the observed $B$-modes are created by lensing. Indeed, while the QE is limited by the cosmic variance of the lensed $B$-modes, the likelihood estimator is able to delens parts of these $B$-modes (provided the noise is below the lensing level of $\sim 5\mu\rm K \, arcmin$), and thus decreases the variance. The MAP lensing mass map estimator essentially achieves what is expected to be the lowest possible reconstruction noise.

The scope of the current paper is to test the performance of the MAP estimator (as implemented in Ref. \cite{Carron:2017mqf}) in the vicinity of galaxy clusters.
Contrary to the likelihood approximations of Refs.~\cite{ Yoo:2008bf, Yoo:2010jd}, or to the MLE of \cite{Lewis:2005fq, Baxter:2014frs, Raghunathan:2017cle}, our MAP estimator does not assume a cluster profile, or the presence of a cluster signal at all, and is thus agnostic in the true deflection field. We take into account the large scale structures (LSS) deflection field together with the one from the cluster specifically. LSS lensing increases the reconstruction noise substantially a low noise levels, making the analysis more realistic.

Our paper is organised as follow. In Section \ref{sec:model} we briefly review the lensing of the CMB by a dark matter halo. In Section~\ref{sec:method} we present the QE and the MAP lensing reconstruction, as well as our matched-filtering to estimate the cluster mass. In Section~\ref{sec:results} we compare the performance of both QE and MAP estimators applied to CMB simulations, and we conclude in Section~\ref{sec:conclusion}.

Our results, including the CMB simulations and the lensing reconstructions, were obtained with the publicly available code LensIt\footnote[1]{\url{https://github.com/carronj/LensIt}}.

\section{CMB lensing by galaxy clusters}
\label{sec:model}

\subsection{NFW profile}

We use the Navarro-Frenk-White profile \cite[NFW,][]{Navarro:1995iw} to model the cluster mass distribution in our analysis. In this model, the density of the galaxy cluster is
\begin{equation}\label{eq:NFW}
    \rho(r) = \begin{cases} \displaystyle 
               \frac{\rho_0}{(\frac{r}{r_s})(1+\frac{r}{r_s})^2} &\quad \text{if} \ r < R_{\text{trunc}} \\
               0  &\quad \text{if} \ r > R_{\text{trunc}}
               \end{cases},
\end{equation}
where $\rho_0$ and $r_s$ are the characteristic cluster density and scale radius respectively. 
We use $M_{200}$ to characterize the mass of the cluster. $M_{200}$ is the mass enclosed in a sphere of radius $R_{200}$, within which the average density of the cluster is $200$ times the critical density of the Universe $\rho_{\text{crit}}$ at the cluster redshift $z$.
We truncate the NFW profile at a radius $R_{\text{trunc}} = 3\times R_{200}$ to obtain realistic mass profiles.
One can write the following relation between $\rho_0$ and $\rho_{\text{crit}}$,
\begin{equation}\label{eq:rho_0}
    \rho_0 = \rho_{\text{crit}}(z) \,  \frac{200}{3} \,  \frac{c_{200}}{ \displaystyle \ln{(1+c_{200})}-\left(\frac{c_{200}}{1+c_{200}}\right)} \; ,
\end{equation}
where $c_{200} = R_{200}/r_s$ is the concentration parameter. It has the following empirical dependence on the cluster mass and redshift \cite{Duffy:2008pz, Geach:2017crt}
\begin{equation}\label{eq:c200}
    c_{200}(M_{200}, z) = 5.71 \, (1+z)^{-0.47} \left(\frac{M_{200}}{2\times10^{12} \, h^{-1}M_{\odot}}\right)^{-0.084} \; .
\end{equation}
Hence, the NFW density profile is quantifiable with these two parameters, the mass\footnote[2]{We express $M_{200}$ in units of solar mass $M_{\odot}$, which is $1.99 \times 10^{30}$ kg.} $M_{200}$ and redshift $z$.

\subsection{Lensing by NFW profile}

The gravitational field created by the mass distribution in the late-time Universe causes deflection of CMB photons. This phenomenon is referred to as weak lensing of the CMB~\cite{Lewis:2006fu}. As a result of weak lensing, the original CMB signal is remapped on the sky. We denote the quantities for the unlensed and lensed CMB as $X (\hat{\mathbf{n}})$ and $\widetilde{X} (\hat{\mathbf{n}})$, respectively. Here, $X$ can represent temperature ($T$) or polarization Stokes parameters ($Q$ and $U$). Assuming the flat-sky approximation, the remapping of the unlensed quantity can be expressed as
\begin{equation}
\widetilde{X} (\hat{\mathbf{n}}) = X (\hat{\mathbf{n}} + \boldsymbol{\alpha}(\hat{\mathbf{n}})),
\end{equation}
where $\boldsymbol{\alpha} (\hat{\mathbf{n}})$ is the deflection field, which can be expressed as the gradient of the lensing potential: $\boldsymbol{\alpha} (\hat{\mathbf{n}}) = \boldsymbol{\nabla}_{\hat{\mathbf{n}}}\phi(\hat{\mathbf{n}})$, neglecting the LSS lensing rotational component (second order in the scalar perturbations). The convergence $\kappa (\hat{\mathbf{n}})$ is the most relevant quantity we work with
\begin{equation}
\kappa(\hat{\mathbf{n}}) = -\frac{1}{2}\nabla^2_{\hat{\mathbf{n}}}\phi(\hat{\mathbf{n}}).
\end{equation}
Since we assume a spherically symmetric profile for the cluster (see Eq.~\ref{eq:NFW}), the cluster convergence is circularly symmetric, and depends only on the radial distance from the cluster center, denoted as $r = |\mathbf{r}|$
\begin{equation}
\kappa_{\text{cl}}(r) = \frac{\Sigma_{\text{cl}}(r)}{\Sigma_{\text{crit}}(z)},
\end{equation}
where $\Sigma_{\text{cl}}(r)$ is the projected surface density of the cluster along the line of sight
\begin{equation}
\Sigma_{\text{cl}}(r) = 2 \int_{r}^{R_{\text{trunc}}}\frac{x\rho(x)}{\sqrt{x^2-R^2}} \d x \; ,
\end{equation}
and $\Sigma_{\text{crit}}(z)$ represents the critical surface density of the universe at the cluster redshift
\begin{equation}
\Sigma_{\text{crit}}(z) = \frac{c^2}{4\pi G}\frac{d_{A,\text{CMB}}}{d_{A,\text{cl}}d_{A,\text{cl-CMB}}} \; ,
\end{equation}
where $c$ is the speed of light, $G$ is the gravitational constant, and $d_{A,\text{CMB}}$, $d_{A,\text{cl}}$, and $d_{A,\text{cl-CMB}}$ represent the angular diameter distances to the CMB, the cluster, and between the cluster and CMB, respectively. 
In the small angle approximation we relate the angular separation $\theta$ and the radial distance from the cluster center with $\theta = \frac{r}{d_{A,\text{cl}}}$.
The expression for the cluster convergence is then
\begin{equation}
    \kappa_{cl} (x) = \frac{2\rho_s r_s}{\Sigma_{crit}(z)}g(x), ,
\end{equation}
where  $x= r / r_s = \theta / \theta_s$ and $g(x)$ is a circularly symmetric function given in Appendix~\ref{A2}. 
We rewrite the cluster convergence as a product of a normalization $\kappa_0$ and a template function $\kappa_t$, which depends on the chosen density profile of the dark-matter halo (and hence on its mass and redshift), so that $\kappa_{\text{cl}} (x) = \kappa_0 \, \kappa_t (x)$.
The normalization $\kappa_0$ is chosen such that $\kappa_t(x=1) = 1$.
Given the definition of $\kappa_0$, it has direct proportionality relation with the mass of the cluster $M_{200}$ \cite{Zubeldia:2019brr},
\begin{equation}\label{eq:tracer}
    \kappa_0 \theta_s^2 \propto \frac{M_{200}}{\Sigma_{\text{crit}}(z)d_{A}^2(z)}
\end{equation}
Hence, $\kappa_0$ works as tracer of the cluster mass and constraining the signal to noise ratio (SNR) for $M_{200}$ boils down to constraining the same quantity for $\kappa_0$, $\displaystyle \frac{\sigma_{\kappa_0}}{\kappa_0} = \frac{\sigma_{M_{200}}}{M_{200}}$.

\section{Cluster mass estimators}
\label{sec:method}
We describe the MAP reconstruction methodology in \ref{sec:estimators}. We follow  Ref.~\cite{Carron:2017mqf} very closely, but had to adapt some details to make the algorithm work reliably at high resolution and high multipoles in the presence of clusters.
In \ref{sec:cluster_mass} we describe the simple matched filter we use to recover the mass of the cluster from the MAP lensing maps.

\subsection{Lensing Reconstruction}
\label{sec:estimators}

We work in the flat sky approximation, and identify multipoles to the plane wave vectors $\vell = (\ell_x, \ell_y)$.
As standard in literature, we use $\vell$ for the CMB multipoles and $\vL$ for the lensing multipoles.   

Let $X$ be the unlensed CMB temperature or polarisation field, $X \in \{T, Q, U\}$, expressed in Fourier space.
The covariance of these unlensed fields are the primordial power spectra 
$\left<X_{\vell}, X_{\vell'}^\dagger \right> = \delta^{\ell}_{\ell'} C_\ell^{\rm unl}$.

The observed CMB field in pixel space can be expressed as 
\begin{equation}
    X^{\rm dat} = BDX + n
\end{equation}
where $D$ is the operator that maps the unlensed CMB modes to the lensed CMB in real space (and thus contains the Fourier transform and lensing remapping operations). The operator $B$ is the linear response matrix of the instrument, including beam and pixel window functions, and $n$ is an independent noise, expressed in pixel space.
When considering the temperature estimator, we have $ X^{\rm dat} = T$, polarization only estimators have $ X^{\rm dat} = (Q, U)$, while the minimum variance estimators have  $X^{\rm dat} = (T, Q, U)$.

For a fixed deflection field the covariance of the observed CMB fields is
\begin{equation}
    \Cov_{\valpha} \equiv \left<X^{\rm dat} X^{\rm dat, \dagger} \right>_{\valpha}  = BDC^{\rm unl} D^\dagger B^\dagger + N \, ,
\end{equation}
where $N$ is the noise covariance matrix in pixel space. 
While if averaging on the deflection fields as well we can write this covariance as 
\begin{equation}
    \Cov \equiv \left<X^{\rm dat} X^{\rm dat, \dagger} \right> = B \mathcal{Y} C^{\rm len} \mathcal{Y}^\dagger B^\dagger + N \, ,
\end{equation}
where $C^{\rm len}$ is the covariance of the lensed CMB fields, given by the fiducial lensed CMB power spectra, and $\mathcal{Y}$ is the Fourier transform operator.

In pixel space, the unnormalized quadratic estimator of the lensing deflection field can be expressed as the product of two filtered CMB maps
\begin{equation}\label{Eq:QE}
    \hat \valpha^{\rm QE}(\vn) = \bar X(\vn) \, \grad X^{\rm WF}(\vn) \, ,
\end{equation}
where the inverse variance filtered and Wiener filtered legs are
\begin{equation}
    \begin{split}
        \bar X &= B^{\dagger} \Cov^{-1} X^{\rm dat} \, , \\
        X^{\rm WF} &= C^{\rm len}  \mathcal{Y}^\dagger B^{\dagger} \Cov^{-1} X^{\rm dat} \, .
    \end{split}
\end{equation}
The normalisation of the QE is chosen to obtain an unbiased estimator, and is expressed in Fourier space as the isotropic response function $\mathcal R^{\rm QE}(L)$~\cite{Hu:2001kj}.
This gives the normalised QE convergence field in Fourier space,
\begin{equation}
    \hat \kappa^{\rm QE}(\vL) = - \frac{1}{2} \frac{i \vL \cdot \hat \valpha^{\rm QE}(\vL)}{\mathcal R^{\rm QE}(L)} \,  \, .
\end{equation}

The maximum a posteriori estimator finds the deflection field that maximizes the likelihood of the lensed CMB fields, assuming a Gaussian prior on the lensing potential, with power spectrum $C_L^{\kappa\kappa, \rm fid}$
\begin{equation}
    \ln \mathcal{P}(\valpha | X^{\rm dat}) = \ln \mathcal{L}( X^{\rm dat} | \valpha) - \frac{1}{2} \, \sum_\vL \frac{\left|\kappa(\vL)\right|^2}{C_L^{\kappa\kappa, \rm fid}}
\end{equation}
where the lensed CMB likelihood is assumed to be Gaussian and given by 
\begin{equation}\label{eq:loglik}
    \ln \mathcal{L}( X^{\rm dat} | \valpha) = -\frac{1}{2} X^{\rm dat, \dagger }\Cov_\valpha^{-1} X^{\rm dat} - \frac{1}{2}\ln \det \Cov_\valpha
\end{equation}
In practice, the MAP lensing deflection field $\hat \valpha^{\rm MAP}$ is found by Newton-Raphson iterations on the posterior, starting from $\hat \valpha^{\rm QE}$ until convergence, using the gradient of Eq.~\ref{eq:loglik} (with respect to the $\valpha$) as search direction. The main term of the gradient can be obtained exactly by running a QE with modified weights on partially delensed CMB maps  \cite[see][for more details]{Carron:2017mqf}. This is the gradient of the first term on the right-hand side in Eq.~\ref{eq:loglik}, quadratic in the data for fixed $\valpha$, whose role is to capture the residual lensing signal. The gradient of the second term on the right-hand side, independent of the data for fixed $\valpha$, is the `mean-field', that removes from the quadratic piece signature of anisotropies unrelated to lensing. In our analysis, there is no masking and the noise is statistically isotropic, making the traditional main sources of mean-field vanish. In principle there is a noise contribution during the iterative procedure: the delensed data noise during the iterative process is not isotropic anymore, because at each step the data is delensed to reduce the CMB statistical anisotropy. This compresses or dilates regions of the sky according to the local estimate of the convergence field, changing the local effective noise levels. This contribution is small and localized at large lensing multipoles, which contributes little to the signal. Hence we neglect the mean field altogether in this paper. 
\begin{figure}
	\centering
	\hspace*{-1.0cm}
	\includegraphics[width=1.\hsize]{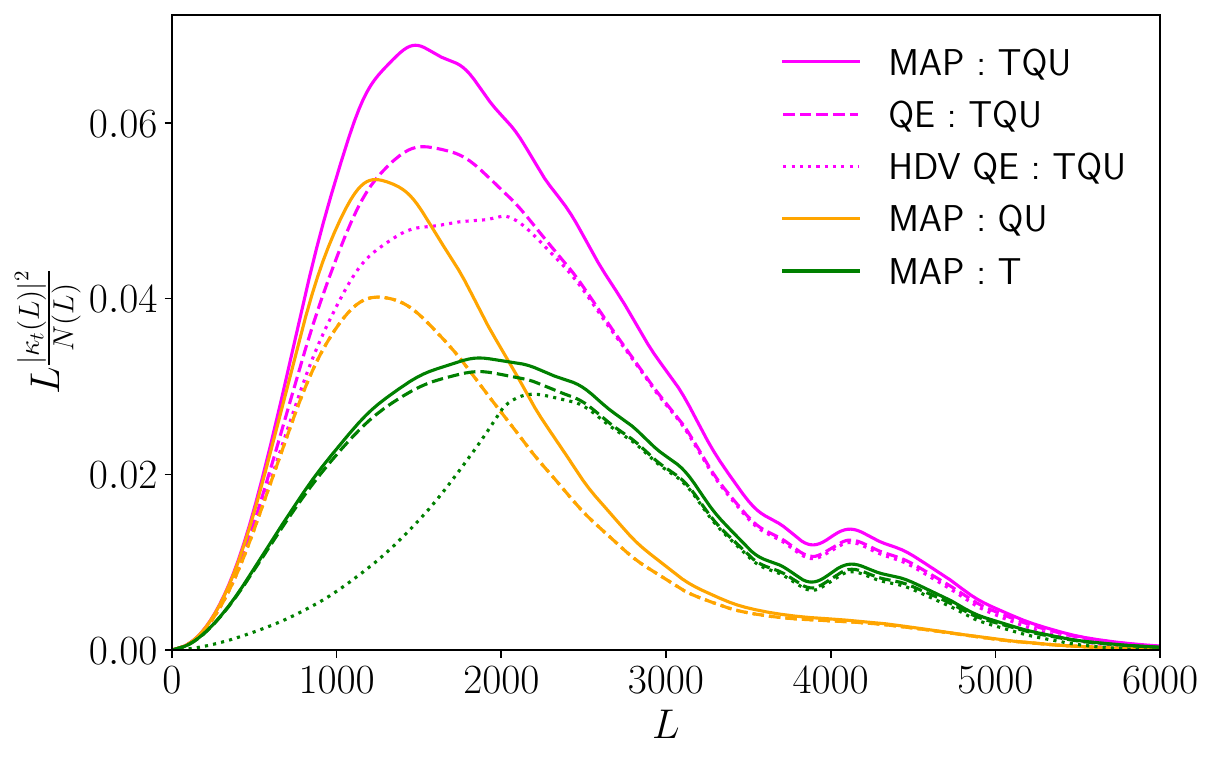}
	\caption{Contribution per lensing multipole to the cluster mass SNR (the integrand of Eq.~\ref{eq:kappa_0_noise}). Green, orange and pink show temperature-only, polarization-only and combined reconstruction respectively. Dashed shows the quadratic estimator case and solid the CMB lensing maximum a posteriori method of this work. Dotted shows the QE with a high-$\ell$ cut on the gradient leg~\cite{Hu:2007bt}. Shown is the case of a $2\times 10^{14} M_\odot$ cluster at $z = 0.7$, for the CMB-S4-like configuration given in the text. While the MAP approach is exactly the same than for optimal CMB lensing power spectrum reconstruction, this is probing much smaller scales (the spectrum SNR is almost entirely confined to $L < 1500$ for this configuration).}
	\label{fig:int}
\end{figure}

Unfortunately, the normalization of this estimator is not tractable analytically. We follow \cite{Legrand:2021qdu,Legrand:2023jne} and obtain an empirical normalization of our MAP estimator from a set of simulations.
We generate 1000 CMB flat sky patches, and lens them with random realizations of the large scale structures deflection field. Note that in these simulations we do not include any cluster signal. The empirical normalization is assumed to be isotropic, and is obtained from the cross correlation between the reconstructed $\hat \kappa^{\rm MAP}$ and the input convergence $\kappa^{\rm in}$, averaged over our set of simulations
\begin{equation}
    \mathcal{W}^{\rm MAP}_L = \left< \frac{C_L^{\kappa^{\rm in}, \hat \kappa^{\rm MAP}}}{C_L^{\kappa^{\rm in}, \kappa^{\rm in}}} \right> \; .
\end{equation}
As shown in \cite{Legrand:2021qdu,Legrand:2023jne}, this empirical normalization is not sensitive to the cosmology, input lensing or data noise, allowing for a correct normalization also when the fiducial ingredients entering the reconstruction process are a poor match to those of the data.

In the posterior maximization process, delensing of the CMB occurs via the operator $D^\dagger$. Our first implementations of the MAP solver in Ref.~\cite{Carron:2017mqf} made the additional assumption of the invertibility of the deflection field, and performed remappings of the maps using a standard bicubic spline algorithm. To improve stability and performance, we updated our flat-sky tools with the lensing method of Ref.~\cite{Reinecke:2023gtp}, based on non-uniform FFTs techniques~\cite{Barnett2019, Barnett2020}. This method is extremely accurate and at the same time removes this unnecessary assumption of an invertible deflection. With this, we found that the search for the MAP point converges without issues on all relevant scales, after only 10-20 iterations.

\subsection{Cluster Signal Reconstruction}
\label{sec:cluster_mass}

Once we have reconstructed the lensing map $\hat{\kappa} (\vL)$ from the simulated CMB data, we fit our theoretical template $\kappa_0\kappa_t (\vL)$.
In practice the template would be obtained with a first guess of the cluster angular scale $\theta_s$. This can be estimated for instance from the tSZ emission of the cluster as in \cite{Zubeldia:2019brr}. Here we take the template which corresponds to the angular scale of the simulated clusters.
Our estimator assumes an isotropic Gaussian noise spectrum $N(\vL)$, such that $\langle{\hat{\kappa} (\vL) \hat{\kappa} (\vL')}\rangle - \langle{\hat{\kappa} (\vL)}\rangle \langle \hat{\kappa} (\vL') \rangle = \delta(\vL - \vL') N (\vL)$. This allows us to construct a minimum variance estimator for $\kappa_0$, given by
\begin{equation}\label{eq:kappa_0}
    \hat{\kappa}_0 = \frac{ \displaystyle \int \frac{\diff^2\vL}{(2\pi)^2} \frac{\kappa_t(\vL) \hat \kappa(\vL)}{N(\vL)}}{\displaystyle  \int\frac{\diff^2\vL}{(2\pi)^2} \frac{|\kappa_t(\vL)|^2}{N(\vL)}}.
\end{equation}
Owing to spherical symmetry, all quantities but $\hat{\kappa} (\vL)$ in the aforementioned expression are functions of $L = |\vL|$. The theoretical error $\sigma^{2,\rm th}_{\kappa_0}$ for the estimator is as follows:
\begin{equation}\label{eq:kappa_0_noise}
\frac{1}{\sigma^{2,\text{th}}_{\kappa_0}} = \frac{1}{2\pi}\int_{L_{\rm min}}^{L_{\rm max}} dL L \frac{|\kappa_t(L)|^2}{N(L)},
\end{equation}
for which we often use the curved-sky expression

\begin{equation}
\frac{1}{\sigma^{2,\text{th}}_{\kappa_0}}=	\sum_{L=L_{\rm min}}^{L_{\rm max}} \left( \frac{2L + 1}{4\pi} \right)\frac{|\kappa_{t,L}|^2}{N(L)}
\end{equation}
In these equations, the noise of the $\hat \kappa(\vL)$ estimate is given by \begin{equation}\label{eq:Noisespec}
 	N(L) = C_L^{\kappa\kappa} + N_L^{(0)} + N_L^{(1)},
 \end{equation}
where $C_L^{\kappa\kappa}$ is the power spectrum of the background convergence profile of the large-scale structure exclusive of the cluster, which contributes to the Gaussian noise on the cluster lensing signal. In the case that $\hat \kappa$ is the QE estimate, $N^{(0)}_L$ is the leading CMB-lensing reconstruction noise (the disconnected four-point function), and $N^{(1)}_L$ the connected part originating from the secondary trispectrum contractions, proportional to $C_L^{\kappa\kappa}$~\cite{Kesden:2003cc,Lewis:2006fu}. In the MAP case, the same Eq.~\eqref{eq:Noisespec} provides a good fit to the spectrum, with both noise terms calculated with partially delensed spectra, following Ref.~\cite{Legrand:2021qdu}. On hypothetical maps where the cluster lensing signal were the only lensing-like signal, only $N^{(0)}_L$ would be present.

On small scales the non diagonal terms in the covariance  $N(L, L')$ of $\hat \kappa(L)$ could become important and reduce the constraining power, as discussed in \cite{Horowitz:2017iql}.
 We checked from our set of simulations (described in the next section) that for all the estimators we considered, and up to $L=5000$, the covariance $N(L, L')$ of $\hat \kappa(L)$ is very close to diagonal, so it is close to optimal to use $N(L)$ for our matched filtering.

In Fig.~\ref{fig:int}, we show the contribution of each lensing multipole to the theoretical signal to noise of Eq.~\ref{eq:kappa_0_noise}. These curves are obtained with the theoretical $N_L^{(0)}$ and $N_L^{(1)}$ of each estimator, considering a  CMB-S4 like experiment with white noise level of $1 \mu \rm{K}$-arcmin in temperature and $\sqrt{2} \mu \rm{K}$-arcmin in polarization and a beam full width at half-maximum (FWHM) of 1 arcmin.
We considered a cluster of mass of $M_{200} = 2\times 10^{14}M_{\odot}$, at $z=0.7$, which matches roughly the expected mean mass and redshift of the CMB-S4 clusters detected with thermal SZ effect \cite{CMB-S4:2016ple}. 

We included there multipoles $\ell_{\text{min}}^{\, \text{CMB}} = 100$ to $\ell_{\text{max}}^{\, \text{CMB}} = 4000$ of the CMB temperature and polarization maps for the lensing reconstruction. 
We compare both the QE and the MAP estimators. We also show the results for a modified QE with a scale cut on the gradient leg of temperature map at $\ell_{\rm cut}=2000$,  noted as Hu-Dedeo-Vale QE (HDV QE) \cite{Hu:2007bt} , which is introduced in more details below.

We see that for both QE and MAP, the lensing scales that dominates the signal to noise are around $L\sim 2000$ in the temperature and around  $L\sim 1100$ in the polarization. The HDV estimator mainly loose information on the large scales, for $L \lesssim 2000$, while smaller scales contain similar information as compared to the QE.
This shows that for this CMB-S4 configuration and cluster convergence profile, temperature and polarization channels are bringing similar level of information, albeit from different scales.  
Finally we clearly see that the MAP estimator outperforms the QE at all lensing scales, and this improvement is mainly due to increased signal from the polarization channel.

\section{Results}
\label{sec:results}
\begin{figure}
	\label{fig:kappa_th}
	\hspace*{-0.5cm}
	\centering
	\includegraphics[width=0.95\hsize]{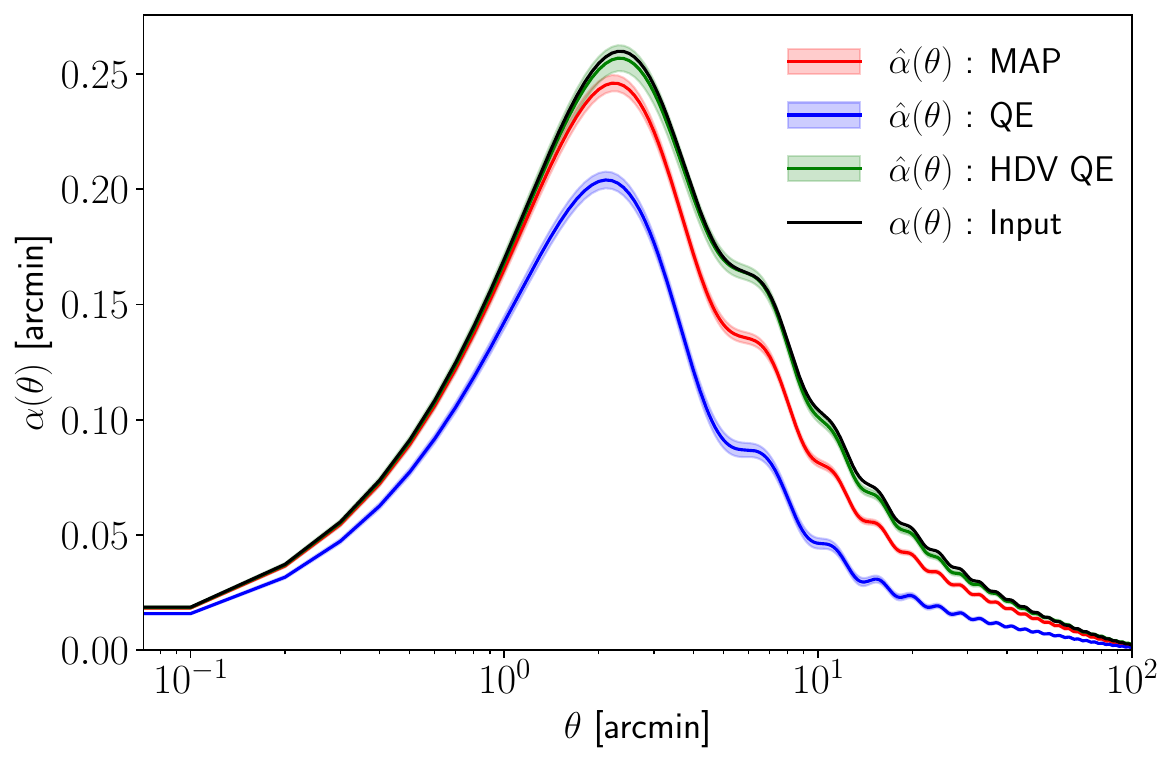}
	\includegraphics[width=0.92\hsize]{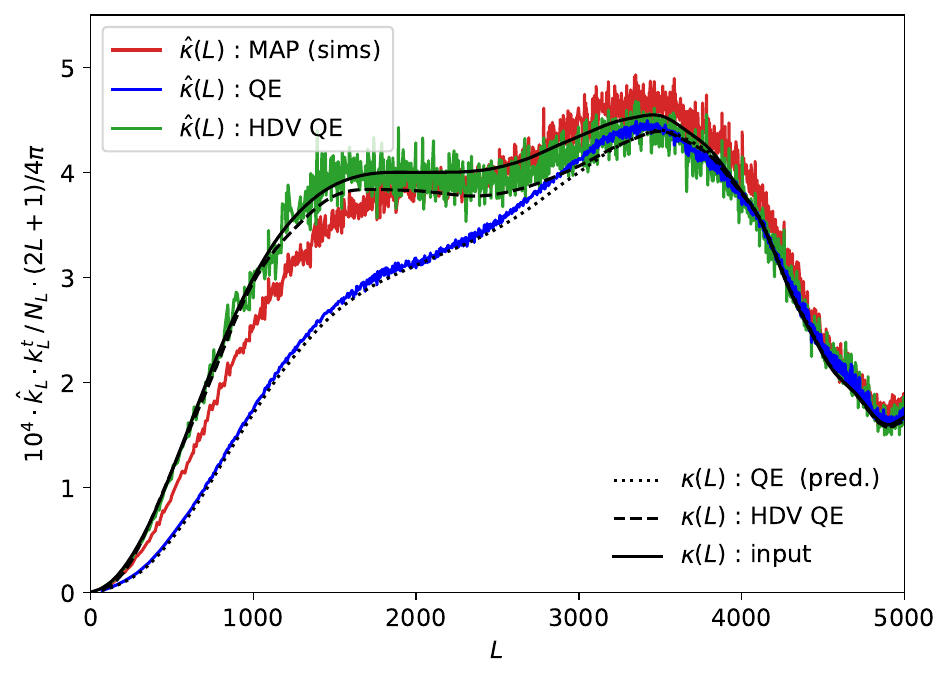}
	\caption{\emph{Top panel}: Reconstructed deflection angle profiles $\alpha(\theta)$ from stacking a set of $1000$ temperature-only reconstructions ($M = 4 \times 10^{14} M_\odot, z=0.7$). CMB multipoles $100 \leq \ell \leq 5000$ are used to reconstruct the lensing multipoles $100 \leq L \leq 5000$. Shown are the standard QE (blue), QE with a high-$\ell$ cut at $2000$ on the gradient leg~\cite{Hu:2007bt} (green, HDV QE), and the maximum a posteriori (MAP) lensing reconstruction of this work (red, without any cuts), and the input truncated to the same $L$ range. The bias in the MAP reconstruction is much reduced compared to the QE but still visible at this cluster mass. In polarization reconstruction no such bias is visible.
		\emph{Lower panel:} The same profiles in harmonic space, weighted at each multipole by their contribution to the cluster mass estimate (see Eq.~\eqref{eq:kappa_0}). An estimate of the map-level reconstruction noise was subtracted in each realization, in order to accelerate convergence, so that the scatter is not representative of an actual data analysis. Analytic predictions are given for the QE and HDV QE cases as the dotted and dashed black lines. The MAP case is slightly tilted in this configuration compared to the input profile (solid black).}
	\label{fig:Bias_sup}
\end{figure}
In this section, we present results on the cluster convergence profile and on the cluster mass from reconstructions of the lensing signal from a set of simulations. 
Our simulations reproduce a CMB S4-like experiment \cite{CMB-S4:2016ple}, with white noise level of  $1 \mu \rm{K}$-arcmin in temperature and $\sqrt{2} \mu \rm{K}$-arcmin in polarization. We assume a beam with a full width at half-maximum (FWHM) of 1 arcmin.
Our flat sky patches have a pixel size of $0.3$ arcmin with 1024 pixels on a side. We simulate the lensing by both the large scale structures and the dark matter halo, assuming (slightly wrongly) they are uncorrelated. The dark matter haloes are in the center of the patches.

We test both the quadratic estimator and the MAP estimator, as introduced in Section~\ref{sec:estimators}, to obtain the estimated $\hat{\kappa}(L)$.
We then estimate the mass of the cluster with the matched filtering, as described in Section~\ref{sec:cluster_mass}. We compare our results for the temperature only (T), the polarization only (QU) and the minimum variance (TQU) estimators.

\subsection{Bias in Temperature QE}

\begin{figure}
	\includegraphics[width=0.95\hsize]{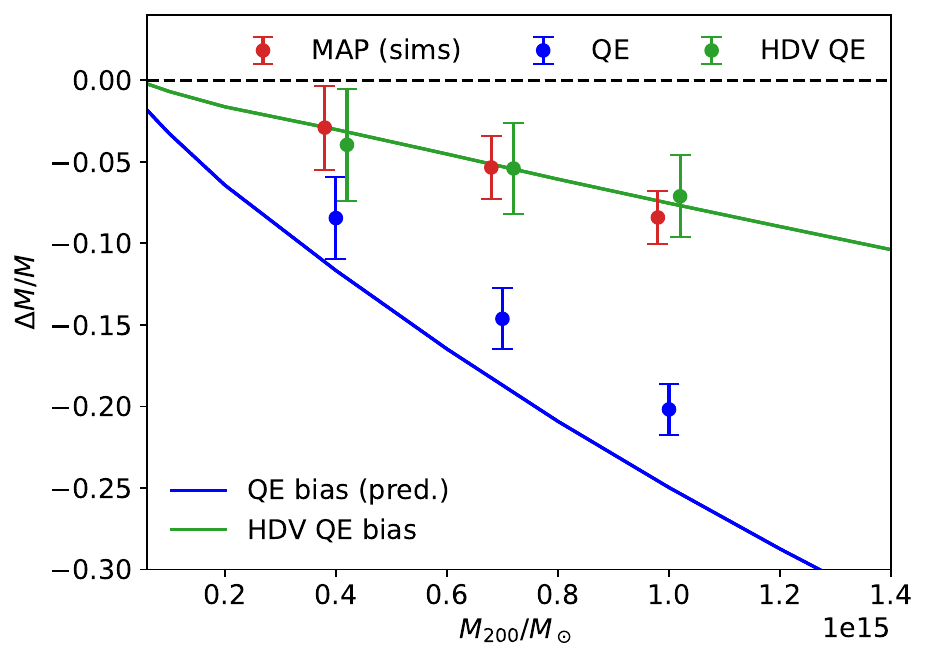}
	\caption{Analytical predictions for the QE-inferred mass bias as a function of halo mass for temperature-only QE lensing mass map reconstruction, with (green, `HDV QE') and without (blue) cut on the gradient leg~\cite{Hu:2007bt}, including CMB multipoles $100 \leq \ell \leq 5000$ and lensing multipoles $100 \leq L \leq 5000$ (here all clusters are set at $z=0.7$). The points show the corresponding results for our simulated reconstructions with $M=4, 7$ and $10\times 10^{14} M_\odot$, together with the result of our MAP method (red). The error bar is our empirical estimate for a sample of a thousand cluster. The bias is a strong function of halo mass and can be very significant for massive haloes. }
	\label{fig:biaspred}
\end{figure}

As discussed in the literature \cite{Maturi:2004zj, Hu:2007bt} the temperature QE is biased low due to strong to moderate lensing close to the cluster center. 
As lensing by the cluster magnifies the background CMB, the gradient of the observed CMB is smaller than the unlensed CMB gradient.
Since the weak lensing signature, i.e. the dipole-like structure in the lensed-unlensed sky, is sensitive to both the strength of the gradient and the mass of the cluster, it decreases that signature as well. 
As a consequence, when we estimate cluster signal using weak lensing temperature QE, the estimator is biased low.

The temperature quadratic estimator is nothing but a product of two filtered maps as given in Eq.~\ref{Eq:QE}. The bias comes mainly from the small scales in the gradient leg. 
A solution to this bias, as demonstrated in \cite{Hu:2007bt}, is to use a low-pass filter for the gradient leg and only include multipoles below $\ell=2000$. We denote this modified QE with a scale cut, as Hu-Dedeo-Vale QE (HDV QE) from now on.
\begin{figure}
	\includegraphics[width=0.95\hsize]{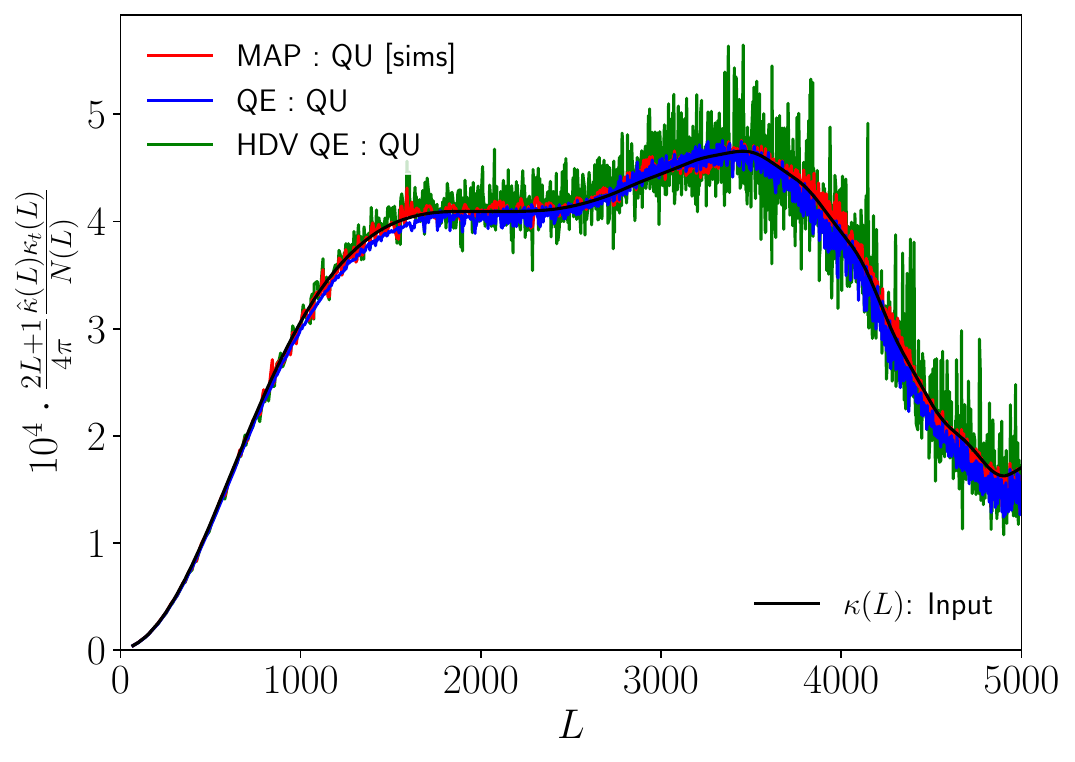}
	\caption{Same as the lower panel of Fig.~\ref{fig:Bias_sup}, for the polarization-only (QU) quadratic and MAP estimators. The QU channel does not suffer from the bias due to strong to moderate lensing close to the cluster center.}
	\label{fig:bias_sup_QU}
\end{figure}

Since the bias is more prominent for massive clusters and small CMB scales, we choose here the clusters parameters to be $M_{200} = 4 \times 10^{14} M_{\odot}, z=0.7$ and use CMB multipoles from $\ell_{\text{min}}^{\, \text{CMB}}=100$ to $\ell_{\text{max}}^{\, \text{CMB}} = 5000$ for the reconstruction.
We run all the estimators (QE, HDV QE and MAP) with the temperature only channel, on $1000$ simulations, and stack the reconstructed maps.
To assess the noise in each reconstruction, we performed additional runs of the estimators on the same simulation, with the same CMB and LSS realization, but without the cluster present. To reduce the variance during the stacking process, for each reconstructed map we first subtract the convergence estimated from the simulation without the cluster, before stacking them.

In Figure~\ref{fig:Bias_sup}, we provide the comparison for the real-space deflection angle profile ($\hat \alpha(\theta)$, upper panel)  and the harmonic space convergence profile ($\hat \kappa(L)$, lower panel). 
The deflection $\alpha$, which is the gradient of the lensing potential $\phi$ is a vector quantity, but only its radial component is non-zero on average, since our clusters are circularly symmetric. 
Hence we only stacked the radial component of $\alpha$.

The lower panel shows
\begin{equation}\label{Eq:Int_kappa0}
	\frac{2L + 1}{4\pi} \frac{\hat \kappa(L) \kappa_t(L)}{N(L)},
\end{equation}
which is the unnormalized contribution of each and every multipole to the mass estimate (empirically, errors are independent to a good approximation). As expected the QE is biased low for such massive clusters, whereas the HDV QE~\cite{Hu:2007bt} does almost unbiased reconstruction of the cluster signal on all scales, at the cost of a 20\% increase in the standard deviation of the mass estimation.
Our iterative estimator in other hand, does get rid of most of the bias, without wasting any information in the small scale regime. For this high cluster mass, our recovered profile still shows some discrepancy to the input, with recovered signal lower than expected on large lensing scales and higher on smaller scales.

On the lower panel, we also plot as the dotted and dashed analytic prediction for the QE bias. 
We do this essentially by brute-force calculation of the QE and HDV QE expectation values, varying CMB and LSS lensing fluctuations, while keeping the cluster lensing deflection fixed, as described in some more details in appendix~\ref{A1}.
The predictions match well, but not exactly, with the empirical findings. 
We could not identify at this moment the origin of the (small) discrepancies. 
One difference is that the analytical predictions are calculated using curved-sky geometry, while the simulations use small boxes in the flat-sky approximation. Fig.~\ref{fig:biaspred} shows the prediction with this tool for the halo mass bias as a function of halo mass, together with our findings on this simulation set. We see that the predicted mass bias for the HDV QE is quite accurate compared to the simulations, while there seems to be a small offset for the QE. It is interesting to note that the MAP bias is very close to the HDV QE bias, while having lower variance.

For comparison, we show the quantity in Eq.~\ref{Eq:Int_kappa0} for the polarization only (QU) estimators in Fig.~\ref{fig:bias_sup_QU}. We see that the bias in the QE, due to the magnification of the background gradient, is almost negligible, contrary to the temperature estimator. We argue that this might come from the fact that the polarization QE is dominated by the EB combination, which is mostly sensitive to the shear and not the magnification. In practice, we also tested that for the EE only QE the bias is smaller than for the TT estimator. 

We now discuss how the iterative estimator also suppresses the noise of the mass estimation compared to QE and HDV QE.
\begin{figure}
	\centering
	\hspace{-1.2cm}
	\includegraphics[width=1.\hsize]{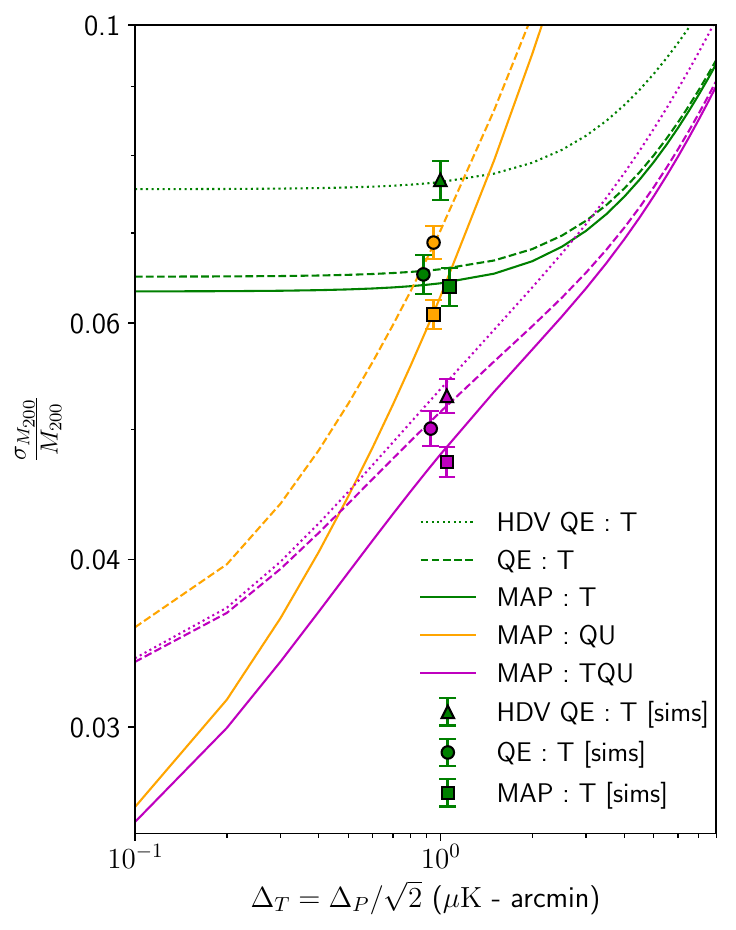}
	\caption{Constraints on cluster mass for a sample of 1000 identical clusters ($M_{200} = 2\times10^{14}M_{\odot}$, at redshift 0.7) as a function of white noise levels. The beam FWHM is $1$ arcmin. Dotted lines (HDV QE) show the QE forecast (Eq.~\ref{eq:kappa_0_noise}) with a cut at $2000$ in the gradient leg~\cite{Hu:2007bt} to remove the QE bias (see Fig.~\ref{fig:Bias_sup}), dashed lines the QE case without any cuts, and solid the maximum a posterior lensing map reconstruction method of this work (MAP). The improvement in constraining power comes from the partial removal from the CMB maps of the lensing signal not directly related to the cluster, and is more prominent in polarization, as expected. Markers with error bars are estimated from our set of simulations, with the x coordinate slightly shifted for clarity. The error bars are obtained from the bootstrapping method (for 1000 clusters), showing good consistency. In the QE case, we applied a simple multiplicative correction to remove the bias in the estimate.}
	\label{fig:forecast}
\end{figure}

\subsection{Cluster mass constraints}

We now consider 1000 simulations with a cluster mass of $M_{200} = 2 \times 10^{14} M_{\odot}$ and $z=0.7$, closer to the expected mean values of the CMB-S4 clusters. We reconstruct the lensing signal with scales between  $\ell_{\text{min}}^{\, \text{CMB}}=100$ and $\ell_{\text{max}}^{\, \text{CMB}} = 4000$, for both polarization and temperature maps.
We reconstruct the lensing convergence fields with the three estimators: QE, HDV QE and MAP, and with the temperature only (T), polarization only (QU) and minimum variance (TQU) estimators. We then run a matched filter on these reconstructed lensing map to estimate the convergence amplitude $\hat \kappa_0$, assuming a template $\kappa_t$ with the characteristic angular size $\theta_s$ corresponding to the input mass and redshift of the cluster profiles.
We use the lensing multipoles between $L_{\text{min}}=100$ and $L_{\text{max}}=6000$ to obtain $\hat \kappa_0$ from Eq.~\ref{eq:kappa_0}. We employ the empirical noise, obtained from 1000 simulations, as the variance $N(L)$ of $\hat \kappa_{L}$ in that equation.

We compute the variance of $\hat \kappa_0$ from our set of 1000 simulations, and compare to the theoretical expectations of the variance from Eq.~\ref{eq:kappa_0_noise}.
Table~\ref{tab:results} summarizes our results.
\begin{table}[]
\centering

\caption{Summary of results on 1000 simulations with input $\kappa_0 = 0.1285$, corresponding to a cluster mass of $M_{200} = 2 \times 10^{14} M_{\odot}$ at $z=0.7$. The table displays the mean $\hat \kappa_0$ evaluated from 1000 simulations along with the corresponding errors. For the T and TQU estimators, HDV QE results are provided, while the QE result is presented for the QU estimator. Additionally, MAP estimated values are shown in the second row. $\sigma_{\kappa_0}^{\text{th}}$ depicts the theoretical error prediction given by Eq.~\ref{eq:kappa_0_noise}. See also Fig.~\ref{fig:forecast}}
    \begin{tabularx}{0.15\textwidth}{|X|X|}
    \hline
    \multicolumn{2}{|c|}{HDV QE : T} \\ \hline
    $\langle\kappa_0 \rangle$      & 0.1276   \\ \hline
    $\sigma_{\kappa_0}$ & 0.0099  \\\hline
    $\sigma_{\kappa_0}^{\text{th}}$ & 0.0096  \\\hline
    \end{tabularx}
    \begin{tabularx}{0.15\textwidth}{|X|X|}
    \hline
    \multicolumn{2}{|c|}{QE : QU} \\ \hline
    $\langle\kappa_0 \rangle$      & 0.1397   \\ \hline
    $\sigma_{\kappa_0}$ & 0.0088  \\\hline
    $\sigma_{\kappa_0}^{\text{th}}$ & 0.0090  \\\hline
    \end{tabularx}
    \begin{tabularx}{0.15\textwidth}{|X|X|}
    \hline
    \multicolumn{2}{|c|}{HDV QE : TQU} \\ \hline
    $\langle\kappa_0 \rangle$      & 0.1337   \\ \hline
    $\sigma_{\kappa_0}$ & 0.0068  \\\hline
    $\sigma_{\kappa_0}^{\text{th}}$ & 0.0067  \\\hline
    \end{tabularx} \\ 
    \hspace*{0.02cm}
    \begin{tabularx}{0.15\textwidth}{|X|X|}
    \hline
    \multicolumn{2}{|c|}{MAP : T} \\ \hline
    $\langle\kappa_0 \rangle$      & 0.1273   \\ \hline
    $\sigma_{\kappa_0}$ & 0.0082  \\\hline
    $\sigma_{\kappa_0}^{\text{th}}$ & 0.0083  \\\hline
    \end{tabularx}
    \begin{tabularx}{0.15\textwidth}{|X|X|}
    \hline
    \multicolumn{2}{|c|}{MAP : QU} \\ \hline
    $\langle\kappa_0 \rangle$      & 0.1342   \\ \hline
    $\sigma_{\kappa_0}$ & 0.0078  \\\hline
    $\sigma_{\kappa_0}^{\text{th}}$ & 0.0081  \\\hline
    \end{tabularx}
    \begin{tabularx}{0.15\textwidth}{|X|X|}
    \hline
    \multicolumn{2}{|c|}{MAP : TQU} \\ \hline
    $\langle\kappa_0 \rangle$      & 0.1331   \\ \hline
    $\sigma_{\kappa_0}$ & 0.0061  \\\hline
    $\sigma_{\kappa_0}^{\text{th}}$ & 0.0062  \\\hline
    \end{tabularx}
    \label{tab:results}
\end{table}
In Fig.~~\ref{fig:forecast} we show the theoretical relative error on the mass for a set of 1000 clusters, as a function of white noise level, for our set of estimators. We also plot the standard deviation we recovered from our simulation set (points with error bars). We employ the bootstrap method, by resampling our simulation set, to estimate the error on this error. 
We see that at the CMB-S4 noise level, the performance of the estimators on simulations are within the expectations. Moreover, we see that the MAP estimator allows to recover the loss in constraining power from the HDV QE estimator.

It is clear that as we go to lower noise level, the polarization estimators eventually dominates the constraints. In this regime, the lensing signal unrelated to the cluster plays an increasingly important role in the error budget. Care must be taken comparing these numbers to the literature, since many works (for example MLE's~\cite{Raghunathan:2017cle}, using Gaussian CMB's generated with lensed CMB spectra instead of a non-Gaussian CMB) do not consider this source of noise.

At the CMB S4 noise level, the MAP estimator detects the cluster signal with an approximate $12\%$, $13\%$ and $20\%$ enhancement in significance compared to HDV QE with TQU, QU and T channels respectively.
While our results are presented based on 1000 simulations, they can be easily extrapolated to the projected number of clusters in CMB S4, estimated to be $10^5$. By employing the MAP estimator for $10^5$ clusters, the mass can be constrained, achieving an accuracy down to 0.47\% for our ideal scenario where no mis-centering or foreground effects are present.

\section{Conclusions}
\label{sec:conclusion}

In this work we demonstrated the use of the maximum a posteriori CMB lensing mass map estimator to reconstruct the mass profile of galaxy clusters.
We showed that the constraints on the cluster masses reach the forecasted ones on simple simulations.
The MAP estimator improves the constraints on the cluster mass by 12\% compared to the HDV QE \cite{Hu:2007bt} with joint temperature and polarization reconstruction, and by 20\% from polarization only. This is due to the absence of scale cuts in temperature, and iterative reduction of the CMB fluctuations sourced by cluster and LSS lensing (in polarization predominantly).

In temperature, the MAP estimator suffers from a much smaller bias than the well-known QE bias, that arises from the misestimate of the background gradient CMB modes. This may be interpreted by the fact that the MAP iteratively estimates the delensed CMB modes, using higher order correlation functions of the CMB maps, to get more accurately the unlensed gradient.
This permits usage of all CMB scales in the reconstruction, without the necessity of scale cuts as in \cite{Hu:2007bt} (provided, of course, that foregrounds or other contaminating signals are under control on the relevant scales).

The MAP reconstruction used here is the exact same that is being developed for generic CMB lensing reconstruction, on wide areas of the sky.  It makes no assumption on what sources the deflections, and a reconstruction on some area of the sky will reconstruct at once all cluster signals in the area in a `optimal',  statistically speaking, manner.
On the practical side, the curved sky MAP implementation received recently considerable increase in speed and accuracy~\cite{Reinecke:2023gtp}, making full-sky analyses perfectly doable in very reasonable time, and converging just as well on realistic, N-body simulations inputs. Results will be presented elsewhere.

In this first work, we used for simplicity flat-sky simulations, each time with one cluster in a small patch. Our simulations went a step further than some other analyses by including the lensing from the background large scale structures together with the cluster. This increases the variance of the QE, and somewhat less for the MAP. We assumed there the background LSS lensing independent from the cluster. In reality, clusters are located at the nodes of the cosmic web, and are highly correlated with overdensity regions. Dark matter N-body simulations such as the Websky \cite{Stein:2020its} and DEMNUni \cite{Carbone:2016nzj} can be used to quantify this. We also neglected many practical issues, such as the impact of mis-centering, which can degrade the constraints on the mass by about $\sim 10\%$ \cite{Madhavacheril:2017onh, Zubeldia:2019brr}. To this and other issues independent from the mass reconstruction process can be applied the same techniques developed for the QEs.

The normalization of the MAP lensing mass map (which is analogous to a Wiener-filter) requires careful handling. Indeed it cannot be obtained accurately analytically and we relied on a set of (cluster-free, with Gaussian input lensing fields) fiducial simulations to obtain it. We showed in \cite{Legrand:2021qdu} that this normalization is independent of the true cosmology, input lensing and data noise statistics of the CMB, hence we expect this procedure to be sufficiently robust in practice.

Much work remains to be done. We did not consider contamination from foregrounds signal, such as the SZ effect, radio point sources or the Cosmic Infrared Background. The thermal SZ effect has a frequency dependence which allows to remove its contribution from the observed CMB maps, at the expense of increased variance. It is possible to reduce the loss of signal in the lensing reconstruction by tSZ cleaning only the gradient leg of the QE \cite{Madhavacheril:2018bxi, DES:2018myw, Patil_2020}. The kinematic SZ however cannot be distinguished from the lensing signal. Efforts have been undertaken to enhance the robustness of MLE-based estimators against tSZ contamination~\cite{Raghunathan:2019tsz, Levy:2023moy}. However, it is expected that the polarization foregrounds are small on the relevant scales, so our MAP estimator forecast in polarization should in principle be robust.

\begin{acknowledgements}
The authors acknowledge helpful discussions with Mathew Madhavacheril and Sebastian Belkner. The computations were performed at University of Geneva on ``Yggdrasil" HPC cluster. SS acknowledges support from the Federal Commission for Scholarships for Foreign Students for the Swiss Government Excellence Scholarship (ESKAS No. 2022.0316) for the academic year 2022-'23. LL and JC acknowledge support from a SNSF Eccellenza Professorial Fellowship (No. 186879).

\end{acknowledgements}
\bibliography{CL}
\appendix
\section{Theoretical Expressions in Clusters Lensing}\label{A2}
The expression for the convergence profile of the galaxy clusters is given by~\cite{Takada:2002qq},
\begin{equation}
    \kappa_{cl} = \frac{2\rho_s r_s}{\Sigma_{crit}(z)}g(x), \text{ where } x=\frac{r}{r_s} = \frac{\theta}{\theta_s}.
\end{equation}
Here $g(x)$ is a circularly symmetric function depending on the truncation of the profile,
\begin{equation}
    g(x) = 
     \begin{cases}
       -\frac{\sqrt{x_{\text{max}}^2-x^2}}{(1-x^2)(1+x_{\text{max}})} + \frac{1}{(1-x^2)^{\frac{3}{2}}}&\cosh^{-1}\left(\frac{x^2+x_{\text{max}}}{x(1+x_{\text{max}})}\right),  \\ &(x < 1)\\
       \frac{\sqrt{x_{\text{max}}^2 - 1}}{3(1+x_{\text{max}})}\left[ 1+\frac{1}{1+x_{\text{max}}} \right],  &(x = 1)\\
       -\frac{\sqrt{x_{\text{max}}^2-x^2}}{(1-x^2)(1+x_{\text{max}})} - \frac{1}{(x^2-1)^{\frac{3}{2}}}&\cosh^{-1}\left(\frac{x^2+x_{\text{max}}}{x(1+x_{\text{max}})}\right),  \\
       &(1< x \leq x_{\text{max}}),\\
     \end{cases}
\end{equation}
where we introduced $x_{\text{max}}$, the ratio of the truncation radius to $r_s$, which in our case is equal to $3\times c_{200}$. 
The analytical form of the Fourier transform to the convergence profile, is given by \cite{Scoccimarro:2000gm, 2011PhRvD..83b3008O, Takada:2002qq}
\begin{equation}\label{eq:ft_anal1}
    \kappa_{a}(l;z) = \frac{M_{200}\Tilde{u}(k=\ell/\chi;z)}{(1+z)^{-2}\Sigma_{\text{crit}}(z)},
\end{equation}
where $\Tilde{u}(k=\ell/\chi;z)$ is given by
\begin{equation}\label{eq:ft_anal2}
	  \begin{split}
    \Tilde{u}(k=\ell/\chi;z) = \frac{1}{m_{\text{nfw}}}[\sin{y}\{\text{Si}[y(1+x_{\text{max}})] -\text{Si}(y)\} \\
    +\cos{y}\{ \text{Ci}[y(1+x_{\text{max}})]-\text{Ci}(y) \} - \frac{\sin{(y x_{\text{max}})}}{y(1+x_{\text{max}})}].
      \end{split}
\end{equation}
Here, we define a new Fourier variable, $k =\ell/\chi$. $\chi$ is the comoving angular diameter distance at that redshift, defined as $\chi(z) = \int_0^z dz'\frac{1}{H(z')} = (1+z)d_A(z)$. In the above equation, $y = (1+z)kr_s$; $\text{Si}(x)$ and $\text{Ci}(x)$ are the sine and cosine integrals and the normalisation $m_{\text{nfw}} = \ln{(1+c_{200})}-\left(\frac{c_{200}}{1+c_{200}}\right)$.

\section{Exact QE bias calculation}\label{A1}
\newcommand{\hn}[0]{\hat n}
In this section we sketch how we obtain the exact expectation value of a quadratic estimator (QE) in the presence of a fixed deflection field. This allows us to predict the mass profiles and mass bias obtained by QE mass estimates. We used a curved-sky implementation, which could certainly made much faster in a satisfactory manner using the flat-sky approximation, but we found it fit enough for purpose.

We consider a generic separable temperature QE, described by two isotropic functions $F_l$ and $G_l$. Following \emph{Planck}-lensing style notation, it may be written in configuration space,
\begin{equation}
    \begin{split}
        _{1}\hat g(\hat n) &= \left( \sum_{l_1m_1} F_{l_1} T_{l_1m_1}\:_sY_{l_1m_1}(\hn) \right) \\
         & \times\left( \sum_{l_2m_2} G_{l_1} T_{l_2m_2}\:_tY_{l_2m_2}(\hn) \right) .
    \end{split}
\end{equation}
In the case of interest, lensing from temperature-only, we have $s =0, t = 1$, and
\begin{equation}
	F_l = \frac{1}{C_l + N_l}, \quad G_l = - \sqrt{l(l+ 1)}\frac{C_l}{C_l + N_l},
\end{equation}
Here, $_{1} \hat g(\hat n)$ is the unnormalized deflection vector (spin-1) field estimate. We want to evaluate this quantity
\begin{equation} \label{eq:avg}
	\av{\: _{1}\hat g(\hat n)}_{\text{fixed $\phi$}},
\end{equation}
which may be used to predict the result of stacking QEs from CMB maps on identical clusters.

Let $\hat n'$ be the undeflected position that corresponds to the observed location $\hn$. Working at fixed $\hn$, by Fourier transforming the $T$-maps, we may write this as a cross-spectrum,
\begin{equation}\label{eq:g}
\begin{split}
	\av{\: _{1}\hat g(\hat n)}_{\text{fixed $\phi$}}
=\sum_{lm} C_l F^*_{lm}(\hn) G_{lm}(\hn),
 \end{split}
\end{equation}
with 
\begin{equation}
\begin{split}
    G_{lm} &= \int d^2n_2 G(\hn, \hn_2) Y^*_{lm}(\hn_2')\\
    F_{lm} &= \int d^2n_1 F(\hn, \hn_1) Y^*_{lm}(\hn_1').
\end{split}
\end{equation}
Here $F$ and $G$ have structure similar to that of a spin-weighted correlation function
\begin{equation}
\begin{split}
	F(\hn, \hn_1) &= \sum_{l_1m_1} F_{l_1} \:_sY_{l_1m_1}(\hn)\:_0Y^*_{l_1m_1}(\hn_1), \\
	G(\hn, \hn_2) &= \sum_{l_2m_2} G_{l_2} \:_tY_{l_2m_2}(\hn)\:_0Y^*_{l_2m_2}(\hn_2). \\
\end{split}
\end{equation}
For an arbitrary deflection field, this is a tough calculation, requiring several spherical harmonics transforms on non-regular grid per each point of interest $\hn$

In the case of cluster lensing things simplify quite a bit owing to
\begin{itemize}
	\item  spherical symmetry, $\av{g(\hat n)}_{\textrm{fixed deflection}}$ will depend on the co-latitude $\theta$ only, and the deflection field does not change the longitude coordinates,
	\item and the fact that clusters are small. With the coordinates such that the cluster lies at the pole, only a small number of $m$'s will be necessary. The circle at latitude $\theta$ has length $\sin \theta$, hence there is an effective $m_{\rm max} \sim l_{\rm max} \sin \theta \sim 5$ for $\theta \sim 10^{'}$ and an analysis with $l_{\rm max} = 5000$. 
	Since it is spin-1, $\av{_1g(\theta)}$ will be heavily dominated by the $m = 1$ component near the pole.
\end{itemize}
We use this to perform the calculation in Eq.~\eqref{eq:avg} by brute force, using the efficient general spherical harmonic transforms of~\cite{Reinecke:2023gtp}.

\end{document}